# iLocater: A Diffraction-limited Doppler Spectrometer for the Large Binocular Telescope

Justin R. Crepp*[a], Jonathan Crass[a], David King[b], Andrew Bechter[a], Eric Bechter[a], Ryan Ketterer[a], Robert Reynolds[c], Philip Hinz[d], Derek Kopon[e], David Cavalieri[a], Louis Fantano[f], Corina Koca[f], Eleanya Onuma[f], Karl Stapelfeldt[f,g], Joseph Thomes[f], Sheila Wall[f], Steven Macenka[g], James McGuire[g], Ronald Korniski[g], Leonard Zugby[g], Joshua Eisner[d], B. Scott Gaudi[h], Fred Hearty[i], Kaitlin Kratter[d], Marc Kuchner[f], Giusi Micela[j], Matthew Nelson[k], Isabella Pagano[l], Andreas Quirrenbach[m], Christian Schwab[n], Michael Skrutskie[k], Alessandro Sozzetti[p], Charles E. Woodward[q], Bo Zhao[r]

[a]Department of Physics, University of Notre Dame, 225 Nieuwland Science Hall, Notre Dame, IN, 46556, US; [b]Institute of Astronomy, University of Cambridge, Madingley Road, Cambridge, CB3 OHA, UK; [c]Large Binocular Telescope Observatory, 933 N. Cherry Ave., Tucson, AZ 85721 US; [d]Steward Observatory, Department of Astronomy, University of Arizona, 933 N. Cherry Avenue, Tucson, AZ 85721, US; [e]Harvard Smithsonian Center for Astrophysics, 60 Garden St., Cambridge, MA 02138, US; [f]NASA GSFC, Greenbelt, MD 20771, US; gJet Propulstion Laboratory, 4800 Oak Grove Drive, Pasadena, CA, 91109, US; [h]Department of Astronomy, The Ohio State University, Columbus, OH 43210, US; [i]405 Davey Laboratory, Pennsylvania State University, University Park PA, US; [j]INAF-Osservatorio Astronomico di Palermo "G. S. Vaiana", Piazza del Parlamento, 1 – 90134 Palermo, Italy; [k]Department of Astronomy, University of Virginia, Charlottesville, VA 22904-4325, US; [l]INAF-Osservatorio Astrofisico di Catania, Via S. Sofia, 78 - 95123 Catania, Italy; [m]Landessternwarte, Zentrum für Astronomie der Universität Heidelberg, Königstuhl 12, 69117 Heidelberg, Germany; [n]Macquarie University, Sydney, NSW 2109, Australia; Australian Astronomical Observatory, Sydney, NSW, Australia; [p]INAF-Osservatorio Astrofisico di Torino, Via Osservatorio 20, I-10025 Pino Torinese, Italy; [q]Minnesota Institute for Astrophysics, School of Physics and Astronomy, University of Minnesota, 116 Church Street, SE, Minneapolis, MN 55455, US; [r]211 Bryant Space Science Center, University of Florida, Gainesville, FL, 32611, US

## ABSTRACT

We are developing a stable and precise spectrograph for the Large Binocular Telescope (LBT) named "iLocater." The instrument comprises three principal components: a cross-dispersed echelle spectrograph that operates in the YJ-bands (0.97-1.30 μm), a fiber-injection acquisition camera system, and a wavelength calibration unit. iLocater will deliver high spectral resolution (R~150,000-240,000) measurements that permit novel studies of stellar and substellar objects in the solar neighborhood including extrasolar planets. Unlike previous planet-finding instruments, which are seeing-limited, iLocater operates at the diffraction limit and uses single mode fibers to eliminate the effects of modal noise entirely. By receiving starlight from two 8.4m diameter telescopes that each use "extreme" adaptive optics (AO), iLocater shows promise to overcome the limitations that prevent existing instruments from generating sub-meter-per-second radial velocity (RV) precision. Although optimized for the characterization of low-mass planets using the Doppler technique, iLocater will also advance areas of research that involve crowded fields, line-blanketing, and weak absorption lines.

*jcrepp@nd.edu

## 1. INTRODUCTION

The Doppler radial velocity (RV) technique continues to drive our understanding of planet masses, their interior structure, and orbital dynamics. Despite early successes involving the discovery of hot-Jupiters and the study of short-period planets initially detected using the transit method, technology advances with the RV technique have reached an impasse. Doppler precision has leveled off and progress is now impeded by the design decisions of the earliest spectrometers.[1] Present-day RV instruments operate under seeing-limited conditions, placing fundamental limitations on their ability to generate high resolution and precision.[2,3,4] As a consequence, existing spectrographs are limited to ~1 m/s single measurement precision, not because of photon noise, but instead as the result of systematic effects such as modal noise and thermal control, as well as the influence of stellar activity which become relevant at the sub-meter-per-second level. This level of performance makes the difference between detecting gas giant planets and Earth-like worlds orbiting in the habitable zone.

We are developing a new type of RV instrument for the Large Binocular Telescope (LBT) named "iLocater," the **i**nfrared **L**arge **Bin**ocular **T**elescope **e**xoplanet **r**econnaissance (iLocater) spectrograph. Unlike seeing-limited designs, iLocater will use adaptive optics (AO) to couple starlight into single-mode fibers (SMFs), in order to generate high spectral resolution (R~150,000-240,000) while eliminating the effects of modal noise. iLocater was initially conceived in 2012 in response to an observatory-wide call for white papers to design second-generation instruments for the LBT. Hardware development is being led from the University of Notre Dame. A scientific and technical collaboration has been established between all seven LBT partner institutions as well as NASA GSFC and NASA JPL. In this paper, we describe the iLocater concept, example science cases, instrument design, and progress to date.

## 2. THE ILOCATER CONCEPT

The ability to implement small (single mode) optical fibers that feed a spectrograph offers the following technical advantages compared to seeing-limited designs:

- higher spectral resolution using a compact optical design;
- elimination of "modal noise" (for both starlight and calibration light);
- elimination of focal ratio degradation and its evolution in time;
- improved imaging quality as a result of slower camera optics;
- two orders of magnitude lower background noise from OH-emission lines and the moon;
- identification of neighboring stars through diffraction-limited observations;
- and improved vacuum levels and thermal control owing to a smaller opto-mechanical footprint.

The requirement to use AO to inject starlight into SMF's well-matches the scientific motivation for developing near-infrared RV spectrographs for exoplanet research.[5,6] Further, observations at red-optical and near-infrared wavelengths are expected to lessen the impact of stellar jitter compared to visible light instruments by effectively reducing contrast between the stellar photosphere and star-spots as well as other surface inhomogeneities.[7] Thus by operating in the near-infrared ($\lambda$=0.97-1.30 $\mu$m) and accepting a well-corrected beam of starlight from both 8.4m mirrors of the LBT, iLocater shows promise to circumvent the obstacles that prevent existing instruments from overcoming the "1 m/s barrier".[8]

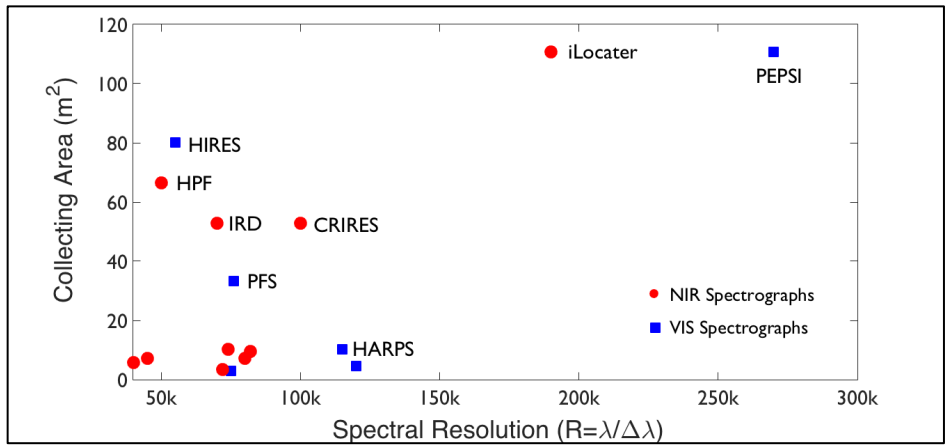

Figure 1. Collecting area versus spectral resolution for existing RV instruments. Infrared and visible spectrographs are denoted by different symbols. Through a unique combination of high spatial resolution and high spectral resolution, iLocater will explore a new parameter space that allows novel programs in the study of stars, brown dwarfs, and extrasolar planets. Note that both iLocater and PEPSI will operate at the LBT: their passbands are complementary with iLocater cutting on ($\lambda$=0.97 $\mu$m) just longwards of where PEPSI cuts off ($\lambda$=0.91 $\mu$m).[9]

## 3. SCIENTIFIC PROGRAMS

iLocater will generate high spatial resolution imaging ($\theta \sim 41$ mas, $1/e^2$ diameter) and high resolution spectroscopy ($\lambda/\Delta\lambda \sim 150,000$-$240,000$) simultaneously. Figure 1 shows a comparison of collecting area (as a proxy for spatial resolution and sensitivity) versus spectral resolution for visible and near-infrared spectrographs installed on large-aperture telescopes. iLocater will open new vistas of exploration in exoplanets, brown dwarfs, and stellar astrophysics by providing the highest resolutions available for any infrared spectrograph from the ground or space. The instrument is designed to reach a single measurement precision of $\sigma=0.6$ m/s ($\sigma=0.4$ m/s goal) including stellar jitter for bright (I<10) M-dwarfs with 30 minutes of integration time using both telescope dishes. This level of performance will provide mass and density estimates for super-Earth's orbiting nearby stars. In addition to stable time-series measurements, for which the spectrograph design is optimized, iLocater will facilitate studies within the solar neighborhood that involve crowded fields, line-blanketing, and weak absorption lines.

### 3.1 Exoplanet Science

iLocater's top scientific priority will be to provide follow-up Doppler measurements for "objects of interest" detected by NASA's TESS space mission.[10] With an average dwell time of 27 days per field, TESS will most effectively explore the habitable zones of nearby K and M stars.[11] These very same targets are ideally suited for RV measurements using a red-optimized RV spectrograph on a large telescope. iLocater will pursue the following research objectives: (i) identify false positive signals through simultaneous AO imaging and spectroscopy; (ii) characterize host star properties ($T_{eff}$, log g, [Fe/H], v sini); (iii) help overcome aliasing effects by measuring planet orbital elements and providing updated ephemerides; and (iv) measure masses and planet densities. It is envisioned that iLocater will focus LBT telescope resources on planet candidates that induce the smallest semi-amplitudes.

In addition to providing essential ground-based support for TESS and any remaining Kepler-2 targets, iLocater will also enable the study of: planets in close-separation binaries (Figure 2); terrestrial planet spin-orbit angles; RV reconnaissance of the youngest transiting planets; exoplanet spin-periods; Doppler tomography; spectra of directly imaged exoplanets; and RV measurements for the TRENDS high-contrast imaging program.[12-19]

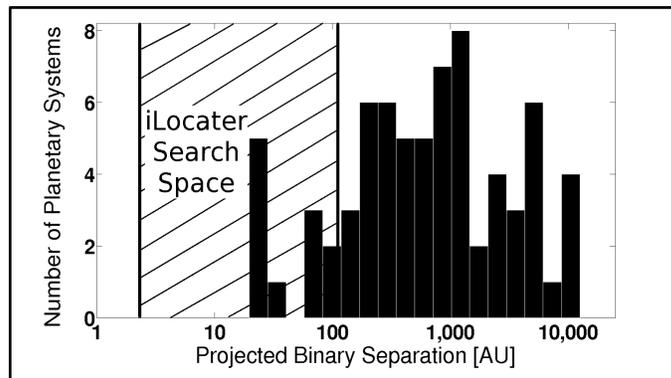

Figure 2. Number of planetary systems discovered orbiting binary stars. Few systems with small projected separations have been searched with the RV technique because existing spectrometers are seeing-limited. iLocater will reduce contamination from neighboring stars permitting Doppler studies of close-separation binaries.

### 3.2 Non-Exoplanet Science

iLocater's high spatial resolution and high spectral resolution measurements will also enable programs in stellar astrophysics including: atmospheric characterization of cold dwarfs; Doppler mapping of substellar objects; kinematic structure of young stellar clusters; characterization of carbon-enhanced metal poor stars; and others.[20,16,21,22]

## 4. INSTRUMENT DESCRIPTION

iLocater is a fiber-fed, cross-dispersed echelle spectrometer that covers the YJ bands at high dispersion ($\lambda/\Delta\lambda \sim 150,000$-$240,000$). Operating in natural guide star mode, it will use the target star as its own reference object for AO correction. Starlight will be injected into single mode optical fibers using the LBT AO system and wavefront sensors within LBTI, an interferometer that gathers light from each telescope.[23] Rather than using LBTI for interferometry, iLocater will use both beams separately to benefit from the equivalent collecting area of an 11.8m telescope.

iLocater comprises three major hardware sub-systems:

- a fiber coupling unit, referred to below as the "acquisition camera," located at the combined focus of LBTI that injects starlight into two SMFs, one for each telescope dish;
- a high-resolution near-infrared (NIR) spectrograph housed within a temperature-controlled cryostat;
- and an etalon comb wavelength calibration system.

The spectrometer will be located away from the telescope's immediate environment to avoid vibrations and minimize temperature changes. A separate room at LBT isolated from disturbances has been identified to house the cryostat. For stability purposes, the instrument will experience a constant gravity vector and have no moving parts. The entire spectrograph will be placed in vacuum. These components are discussed in the following sections.

### 4.1. Wavelength Range

Most pronounced in the K-band, telluric absorption lines have limited previous attempts at precise NIR RV measurements.[24] Efforts to circumvent this problem generally involve wholesale removal of spectral features using software, but this approach severely limits the amount of RV information extracted from remaining starlight. In the K-band telluric absorption results in ~80% loss of spectral data.[25] Fortunately, there exists a NIR band relatively free of atmospheric absorption. The Y-band is clean compared to other infrared windows and also provides access to deep stellar lines rich in Doppler information, especially with K-dwarfs and M-dwarfs.[26,27] iLocater will operate in the YJ-bands, a region that avoids the thick forest of tellurics found in K-band and is also not riddled with bright OH-lines as found in H-band.[28]

### 4.2. Acquisition Camera

iLocater's acquisition camera will be installed at the combined focus of LBTI where two f/41.2 beams combine to form an f/15 envelope.[23] The acquisition camera comprises two optically identical subsystems since differing incident beam angles at the LBTI focus precludes efficient coupling into separate fibers using a common optical path. LBTI provides an excellent environment for SMF coupling since the facility already requires and maintains a stable environment for interferometry. Further, a dual-aperture mounting structure allows the sky to rotate around on-axis sources, as with angular differential imaging, leading to further stability compared to alternative configurations.

The LBT "extreme" AO system uses a pyramid wavefront sensor and has demonstrated impressive correction at short wavelengths.[29,30] For example, Fig. 3 (left panel) shows an image of the star HIP 48455 (V=3.85) taken with one telescope dish. LBT AO delivered 35% Strehl and cleared out a "dark hole" at $\lambda=0.63$ μm.[31] These results show great promise for iLocater, which will operate at longer wavelengths (see Fig. 3, right panel).

Optical design of the fiber-injection camera system is currently underway and closely follows the same principles outlined in detail in Bechter 2015. iLocater will use (germanium-doped) Fibercore SM980 (5.8/125) SMFs which provide a mode-field diameter (MFD) = 5.8 μm at $\lambda=980$ nm and numerical aperture (NA) = 0.14. The AO corrected beam from each telescope is collimated, passes through a zero beam deviation atmospheric dispersion corrector (ADC) and is then split using a dichroic into separate image monitoring and fiber-coupling channels. The imaging channel magnifies the PSF to allow precise centroiding of the incident PSF to monitor beam position and quality. The fiber channel uses a pair of lenses to demagnify the telescope PSF diameter to best match that of the fiber MFD while not exceeding the fiber acceptance NA.[32,33]

The above approach was successfully demonstrated on sky at the LBT in April 2016 where Y-band light was injected into SMFs with sustained coupling efficiencies greater than ~20% (Bechter et al. 2016). A range of visual magnitudes (V=4-11), airmasses, and seeing conditions were explored (Crass et al. 2016, in prep.). Efficiencies of 2-5 times higher than that required for TESS targets were consistently achieved. The planned LBT AO "SOUL" upgrade, which will increase the wavefront sensor frame rate from 1.0 kHz to 2.0 kHz, is expected to improve SMF coupling efficiencies by an additional ~50-70%.[34]

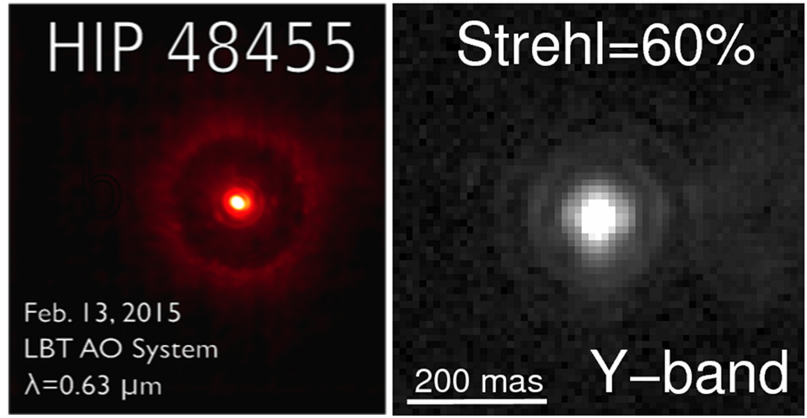

Figure 3. (Left) On-sky demonstration of 35% Strehl at visible wavelengths including correction for non-common-path errors using the LBT "extreme" AO system.[31] Data was taken in February 2015 (sec(z)=1.05) under seeing conditions of θ=0.8" with the AO system running at f=990 Hz. (Right) On-sky demonstration of high Strehl observations in the Y-band at the LBT on Sept. 16, 2014. The star, HD 252468, has an apparent magnitude (R=8.5) comparable to that of a bright TESS target. Several Airy rings can be seen. The FWHM is 30 mas. This figure was reproduced with permission of the INAF team who are developing a visible light imager for the LBT named V-SHARK.

### 4.3. Spectrometer Optical Design

iLocater's optical design requires high spectral resolution (R>150,000) to measure absorption line asymmetries as well as diffraction-limited performance across the YJ-bands (λ=0.97-1.30 µm). As the instrument target operating temperature is T~58 K (Section 4.6), an all-reflecting rather than transmissive design was pursued to establish uniformity of materials selection and to avoid differential motion of compound lenses within their mounts (further the thermal properties of many optical glasses at cryogenic temperatures are not well understood). Each of the spectrometer optics will be gold coated (λ/20 surface quality) Zerodur to maximize throughput and minimize distortions when cooling down from room temperature.

Three fibers, each separated by 125 µm, inject light into the spectrograph: one from each LBT primary mirror and a third for calibration and sky-background subtraction. The fibers are separated in the cross-dispersion direction and provide an ultra-stable Gaussian beam that eliminates the effects of modal noise. High spectral resolution is achieved as a consequence of the small input image set by the SMF core diameter. This feature reduces the required beam diameter (~25 mm) and hence the size of individual mirrors (50 mm – 160 mm diameter); it also translates into a much smaller overall opto-mechanical footprint (50 mm × 500 mm) compared to conventional seeing-limited spectrographs where the fiber is an order of magnitude larger. Figure 4 illustrates how SMFs reduce OH-emission to negligible levels.

We use both OpticStudio Zemax and Code V to model and optimize the spectrograph. A lay-out of the optical train is shown in Figure 5. The design can be broken down into three modules:

- a magnifying relay that slows the f/3.6 output beam from the fibers to f/12.5
- a collimation and dispersing stage
- a reimaging camera to collect the spectrum

Magnification of the fiber input beams is achieved using a three-mirror relay. The first mirror surface is spherical while the second and third surfaces are conics. The resultant f/12.5 beam is directed into the collimating stage using a flat fold mirror. The beam is then collimated using an f=250mm off-axis paraboloid (OAP1), and directed towards a Richardson (Newport) echelle grating (13.33 lines/mm) blazed at an angle of tan(80.7)=6.1. The beam diameter as incident onto this grating is d=24.7 mm. The pupil is at the mid-position of the grating surface, and the grating is rotated by ~1.7 degrees to separate the input and output beams, i.e. the system is in near-Littrow configuration. The dispersed beam is redirected back towards OAP1, and then towards the "spectrum mirror," so called because the dispersed image plane is formed close to its surface.

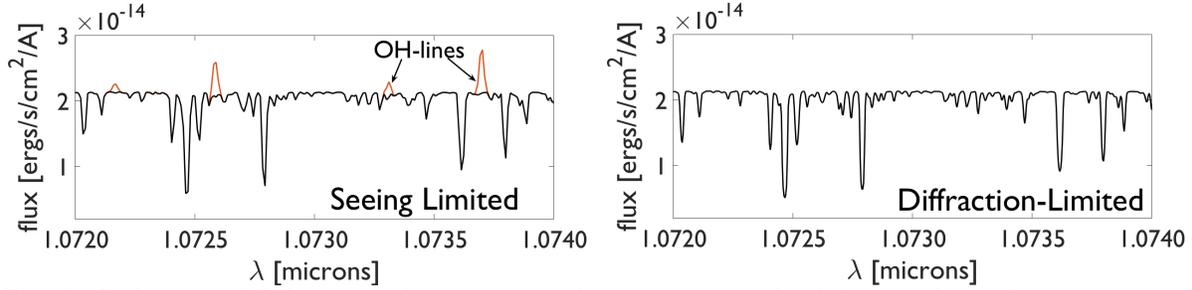

Fig. 4. Reduction of OH-emission line intensity when operating at the diffraction limit. Owing to a much smaller PSF size, iLocater's SMFs will reduce the intensity of OH-lines by 2-3 orders of magnitude compared to seeing-limited instruments. Shown to scale is the spectrum of a V=12, M0V star (black) compared to sky emission lines (orange) in the Y-band. Lines lost to OH emission would be reduced to only ~2% as a result of this effect.

The beam is then directed towards OAP2 (f=250 mm), which collimates the dispersed beam, and forms a pupil image on the surface of a Richardson R53-640R reflection grating blazed at $\lambda=1.2$ μm that cross-disperses the spectra. Finally, a three-mirror-anastigmat (TMA) reimages the resultant spectra onto an H4RG-10 infrared detector array (4096×4096, 10μm pixel size) from Teledyne. As with the magnifying relay, the mirrors in the camera TMA comprise one spherical mirror and two with conical surfaces. Given the need to magnify the fiber as reimaged onto the detector, iLocater's camera has a slow (f/16) focal ratio that provides excellent imaging quality.

Following the technique described in Robertson & Bland-Hawthorn 2012, we over-size optical components by a factor of two (compared to the $1/e^2$ Gaussian width) to avoid diffractive "ringing" (Airy features) when truncating the Gaussian input beam. Rigorous physical optics propagation (POP) simulations show that we reach a spectral resolution of R=150,000–240,000 with ~3.1 pixel sampling as averaged across the YJ-bands. A total of 39 spectral orders (orders 114 – 152) covering the wavelength range $\lambda=0.97 – 1.30$ μm are reimaged onto the detector. Dispersive power changes significantly along the spectral orders due to the extreme R6.1 grating angle. Figure 6 shows iLocater's detector format. The lower orders overfill the detector leading to ~15% loss of spectral information in the J-band while collecting all of the Y-band up to $\lambda=1.09$ μm. Figure 7 shows a spot diagram for the perfectly aligned system illustrating that it is diffraction-limited across the array. Figure 8 displays a theoretical representative spectrum with ~0.02 Angstrom samples across one order of the Y-band. Ultra-high resolution will aid in measuring Doppler shifts and identifying line asymmetries, a signature of stellar activity. Spectrograph design parameters are listed in Table 1.

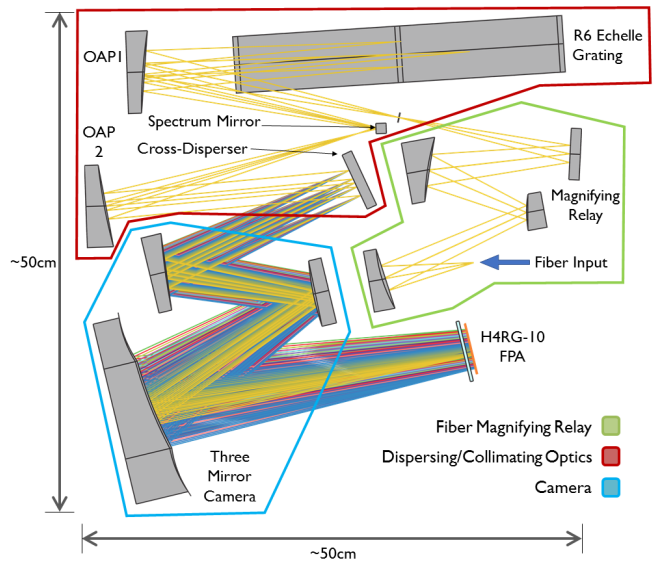

Figure 5. Zemax optical model. iLocater will have a small opto-mechanical footprint leading to improved optical quality, vacuum levels, and thermal control compared to seeing-limited instruments.

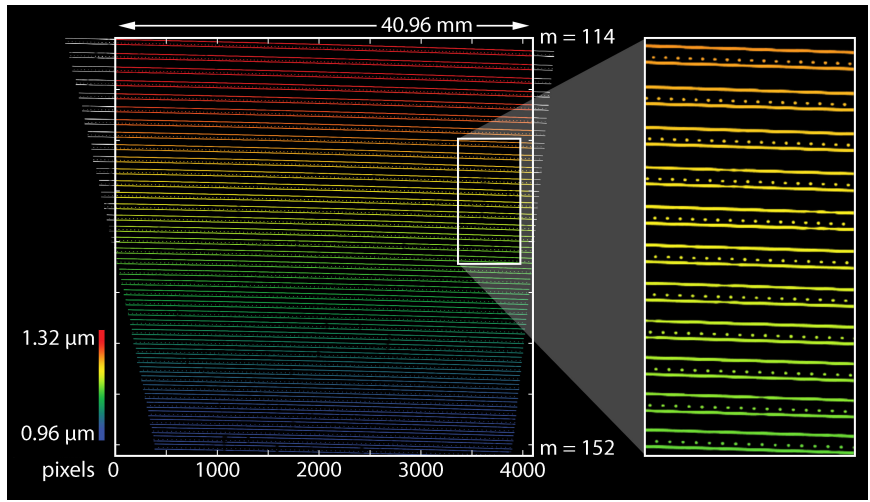

Figure 6. iLocater's spectral format showing all 39 orders and three fiber spectral traces. An ultra-compact design generates diffraction-limited performance across the λ=0.97-1.30 μm range. An H4RG (4096 × 4096 pixels) captures the Y-band and 85% of J-band.

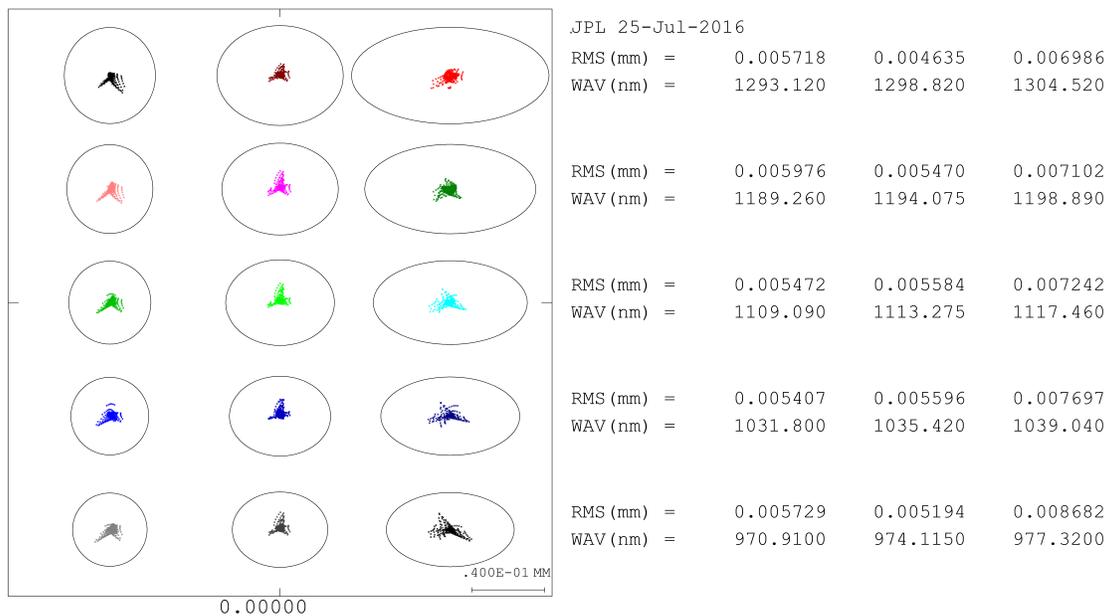

Figure 7. Code-V optimization results showing spot diagrams from orders 114, 124, 133, 143, and 152 to quantify imaging quality as it varies across the H4RG array. iLocater's optical design delivers diffraction-limited performance across the Y and J-bands leading to high spectral resolution.

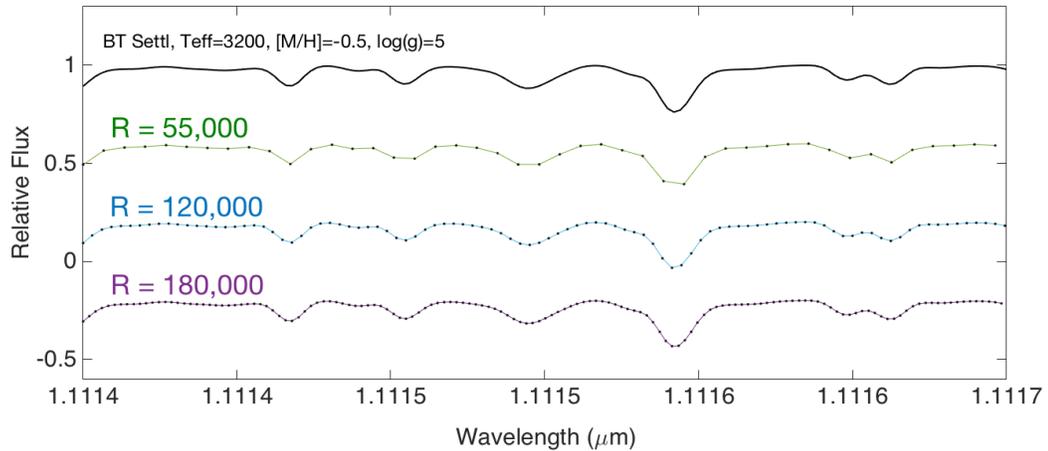

Figure 8. The effects of resolution on spectral sampling. For clarity, model spectra have not been convolved with an instrument response function nor Doppler broadened due to rotation. In addition to improved velocity precision, spectral resolutions greater than R~150,000 are needed to permit absorption line shape measurements that help discriminate between false positive signals caused by stellar activity and orbiting planets (Fischer et al. 2016). iLocater will assess line asymmetries by generating a mean resolution of R~180,000 with 3 pixel sampling per resolution element.

## 4.4. Detector Selection

iLocater's initial optical design included only the Y-band with two H2RG's joined in a 2k x 4k mosaic. However, unavoidable gaps (3.6 mm wide) between arrays resulted in unacceptable losses in wavelength coverage (see JWST NIRCam reference document). An H4RG-10 circumvents this problem and permits the use of smaller pixels (10 μm pitch), thus a more compact camera design. Further, inclusion of the J-band in the cross-dispersion direction provides access to more stellar absorption features. Table 2 shows a comparison of detector specifications for various instruments. The H4RG-10 is competitive in each category including read-out noise (RON), dark current (DC), and quantum efficiency. Both Subaru/IRD and CFHT/SPIRou will use H4RG's.[35,36]

| Parameter | Value | Units | Notes |
|---|---|---|---|
| Spectral Resolution | 150k-240k | unitless | range across YJ bands |
| Sampling (FWHM) | 2.7-3.3 | pixels | average across YJ bands |
| Bandpass | 0.97 - 1.30 | microns | YJ bands |
| Orders | 114-152 | unitless | YJ bands |
| Beam Diameter | 24.7 | mm | $1/e^2$ diameter |
| Grating Blaze Angle | 80.7 | degrees | R6.1 from Newport |
| Grating Line Density | 13.33 | mm$^{-1}$ | 75 μm groove spacing |
| X-Disp. Line Density | 250 | mm$^{-1}$ | 4 μm groove spacing |
| Camera EFL | 400 | mm | three-mirror anastigmat |
| Detector Array | $4096^2$ | pixels | cross-dispersed format |
| Pixel Size | 10 | microns | H4RG from Teledyne |
| Fiber MFD | 5.8 | microns | evaluated at λ=0.98 μm |
| Fiber Angular Size | 41 | mas | projected $1/e^2$ diameter |
| Fiber NA | 0.14 | unitless | Fibercore SM980 |
| Limiting Magnitude | V=16 | mag | M4V, S/N=10 pix$^{-1}$, 30 min. |

Table 1. Spectrograph design parameters.

| Instrument | Detector | Passband | Npix | RON (e-) | DC (e- / s) | QE (%) |
|---|---|---|---|---|---|---|
| HARPS | E2V 44-82 | Visible | 4096 × 4096 | 2.9 | 0.5 | 80% |
| HIRES | MIT/Lincoln | Visible | 3 × (2048 × 4096) | 3 | 0.0006 | 65% |
| CRIRES | Aladdin III InSb | NIR | 4 × (512 × 4096) | 10 | 0.05 | 80% |
| iLocater | H4RG-10 | NIR | 4096 × 4096 | 12 | 0.0023 | 90% |

Table 2. Comparison of CCD's versus NIR arrays for existing Doppler instruments.

## 4.5. Wavelength Calibration

Calibration lamps used by the previous generation of spectrometers, such as Th-Ar and iodine gas cells, must be replaced to achieve ultra-precise RV measurements.[8] iLocater will use an etalon to illuminate the spectrometer with a dense set of uniformly bright and stable lines. An inexpensive version of a frequency comb, stabilized etalons may be used to track instrument drifts at the level of 1 part in $10^{11}$.[2] Gurevich et al. 2014 have demonstrated RV precisions of ~3 cm/s when the system is locked to the hyperfine transition lines of a Rb vapor cell.[37] Etalon light will be injected into the spectrometer via the acquisition camera system to eliminate non-common-path errors. Calibration data will generally be obtained in between exposures during telescope slews although a contemporaneous mode is also available. A U-Ne lamp will be used to confirm near-infrared wavelength solutions.[38]

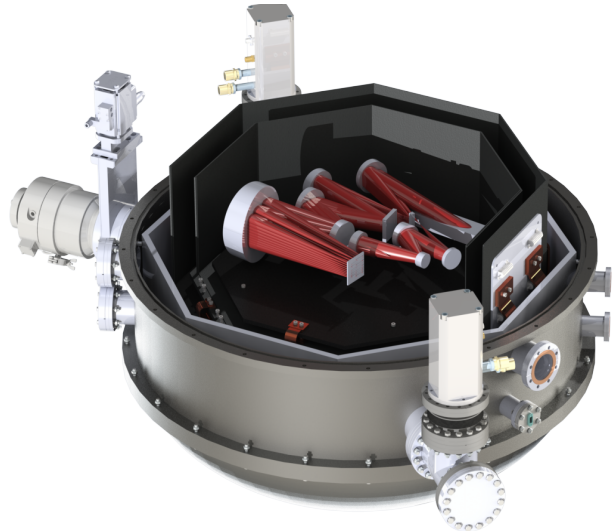

Figure 9. CAD rendering of the iLocater cryostat showing spectrograph optics located inside of a nested set of radiation shields and stainless steel vacuum jacket.

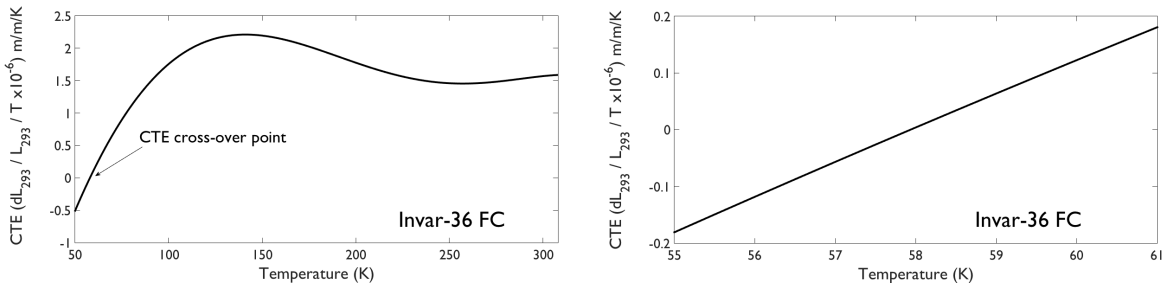

Figure 10. (Left) Coefficient of thermal expansion (CTE) of Invar-36 from Munoz & He 2006.[41] (Right) Zoomed in portion of same curve showing a cross-over point where the CTE crosses zero. iLocater's optical breadboard and mounts will be driven to a temperature of T=58±1 K, and held to better than δT =1 mK rms, in order to reduce expansion and contractions to CTE < 1 x $10^{-7}$ m/m/K. This level of stability corresponds to Doppler RV shifts of only δV~c δT * CTE ~ 3 cm/s.

## 4.6. Cryostat and Thermal Control

iLocater's cryostat must create an intrinsically stable environment for the optical components. A design requirement has been set such that thermal perturbations contribute at most $\sigma=0.1$ m/s to the overall RV error budget. This level of precision corresponds to sub-milliKelvin control of the optics and breadboard. Due to the wavelength cutoff of the detector ($\lambda=2.5$ µm), it is also necessary to operate the instrument at cryogenic temperatures to reduce thermal background. iLocater's cryostat and thermal control system are designed to achieve these requirements.

The spectrograph is housed within a stainless steel vacuum chamber ($P < 10^{-7}$ Torr) that is robust to pressure cycles, helps to damp vibrations, and reduces thermal convective transport to negligible levels. The cryostat will use active thermal control in combination with radiative shielding (Figure 9). Vacuum pressure is passively maintained using charcoal getters. Multi-layer insulation (MLI) reduces radiative loads from the vacuum chamber walls onto the instrument radiation shields. The outer radiation shield is actively controlled to ~1 mK using a closed-loop temperature control system originally designed at the University of Virginia and implemented with the APOGEE instrument at Sloan.[39] A similar system is also being developed for the Habitable-zone Planet Finder (HPF), which has demonstrated sub-mK stability over several weeks.[40] Calibrated semi-conductor diodes provide temperature measurements and work in concert with precisely controlled resistive heaters. An inner radiation shield further dampens temperature fluctuations. See Crass et al. 2016 for further details.

Due to the small optical footprint of the spectrograph, it is possible to afford more intrinsically stable materials for the construction of iLocater compared to existing Doppler instruments. The instrument optical board and opto-mechanics will be made from Invar-36 (36% Nickel alloy), which has a coefficient of thermal expansion (CTE) that is an order of magnitude smaller than Aluminum 6061. Although originally invented for room temperature operation, Invar has a CTE value of zero at T=58K (Figure 10), such that there is a transition from expansion to contraction at this temperature (Munoz & He 2006). This feature sets the operating temperature of the instrument.

Rather than using a liquid cryogens tank, which requires regular fills that thermally "shock" the system (also discouraged by the observatory), iLocater will employ a pair of closed-cycle cryocoolers (PT-60 from Cryomech) to avoid thermal transients while also greatly simplifying the overall design. The compressors will be remotely located to minimize vibrations in the local environment of the chamber while the cold heads will be independently mounted from the chamber wall and connected by vacuum baffles. The cold-tip of each cryocooler will be connected using flexible copper straps to a copper annulus mounted at the base of the instrument which provides thermal connections to the outer shield. Similar experiments have reduced mechanical disturbances to levels comparable to that of liquid nitrogen boil off.[1]

Testing of the thermal control system is being undertaken at Notre Dame using a laboratory vacuum chamber (Figure 11). This test system will help to validate iLocater's thermal control strategy both at room temperature and operating temperature with a planned system upgrade to include a cryocooler. It will also be used to assess feed-through strategies, measure outgassing rates, and measure thermal background levels.

## 5. SUMMARY

Existing RV instruments operate under seeing-limited conditions placing fundamental limitations on their ability to measure precise Doppler shifts. As a consequence, these instruments are limited to ~1 m/s precision and will therefore require hundreds, if not thousands, of measurements to precisely determine the mass of a single TESS planet that may be located near the habitable zone. iLocater is a new type of spectrograph being developed for the LBT that will operate at the diffraction limit and use single mode fibers. As a consequence of this design feature, a number of "tall-poles" common to the error budgets of present-day Doppler instruments will be improved by orders of magnitude (background noise, thermal stability, vacuum pressure, imaging quality) and in some cases eliminated entirely (modal noise). Rather than implementing fiber scrambling, pupil slicing, and enormous gratings, etc., iLocater will use small optical fibers to achieve an opto-mechanical footprint that lends itself to an ultra-stable environment. Driven by the desire to perform follow-up measurements for NASA's TESS mission, we anticipate that iLocater will be delivered to the LBT as TESS science operations begin to ramp up and the first habitable-zone worlds are being detected around nearby stars in 2018.

---

[1] The "cryo-cooler vibrations problem" has been addressed by several NASA space flight instruments.[42] Damping factors of 100x are regularly obtained and show promise to meet iLocater's stringent vibration amplitude requirement.

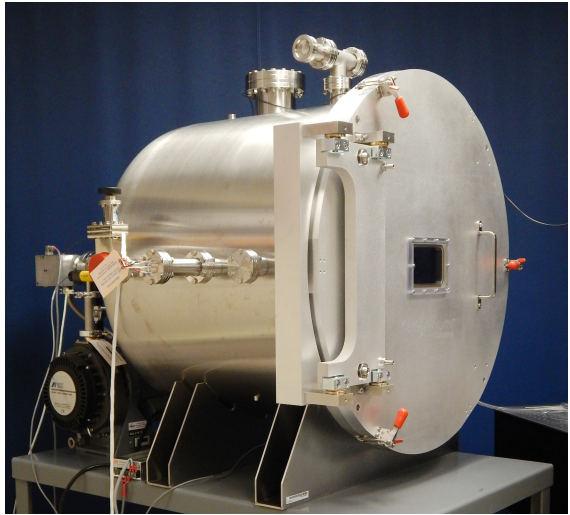

Figure 11. Experimental vacuum chamber undergoing tests at Notre Dame.

By virtue of its "direct to single mode" operation, iLocater will become the first spectrometer optimized for Doppler measurements that successfully decouples the influence of the telescope from the instrument spectral resolution and response. iLocater designs may therefore be effectively "cloned" for other telescopes to facilitate the rapid development and deployment of powerful, yet affordable, RV instruments at other observatories that broadly benefit the exoplanet community. Such designs will be essential for the next generation of telescopes with 20-30m diameters for which seeing-limited spectrographs may prove prohibitively expensive.

**Acknowledgements.** We are grateful for the help of LBTO staff and the INAF V-SHARK team, especially Fernando Pedichini, for assistance setting up fiber coupling experiments with LBTI. We thank the many undergraduates who have worked on the iLocater project including John "Jack" Brooks, Michael Foley, Edward Kielb, Elliott Runburg, Alyssa Runyon, and David Shaw. J. Crepp acknowledges support from the NASA Early Career Fellowship program to develop a fiber-coupling demonstration unit for the LBT. The iLocater team is also grateful for contributions from the Potenziani family and the Wolfe family for their vision and generosity.

The LBT is an international collaboration among institutions in the United States, Italy and Germany. LBT Corporation partners are: The University of Arizona on behalf of the Arizona university system; Istituto Nazionale di Astrofisica, Italy; LBT Beteiligungsgesellschaft, Germany, representing the Max-Planck Society, the Astrophysical Institute Potsdam, and Heidelberg University; The Ohio State University, and The Research Corporation, on behalf of The University of Notre Dame, University of Minnesota and University of Virginia.

A portion of the research in this paper was carried out at the Jet Propulsion Laboratory, California Institute of Technology, under a contract with the National Aeronautics and Space Administration (NASA). T